\documentclass[prd,showpacs,nofootinbib,preprint,10 pt]{revtex4}
\usepackage{amsmath,graphicx,color,xcolor,epsfig}
\usepackage{epstopdf}
\usepackage{hyperref}

\begin{document}
\begin{titlepage}

\begin{center}

{\bf\Large\boldmath Spatial dependence of entanglement renormalization in $XY$ model}\\[15mm]
\setlength {\baselineskip}{0.2in}
{\large  M. Usman, Asif Ilyas, Khalid Khan}\\[5mm]

{\it Department of Physics, Quaid-i-Azam University, Islamabad 45320, Pakistan.}\\[5mm]

\end{center}

{\bf Abstract}\\[5mm] 
\setlength{\baselineskip}{0.2in} 
In this article a comparative study of the renormalization of entanglement in
one, two and three dimensional space and its relation with quantum phase transition (QPT) 
near the critical point is presented by implementing the Quantum Renormalization Group (QRG)
technique. Adopting the Kadanoff's block approach, numerical results for the
concurrence are obtained for the spin -1/2 $XY$ model in all the spatial
dimensions. The results show similar qualitative behavior as we move from
the lower to the higher dimensions in space but the number of iterations
reduces for achieving the QPT in the thermodynamic limit. We find that in the two
dimensional and three dimensional spin -1/2 $XY$ model, maximum values of the
concurrence reduce by the factor of $1/n$ $(n=2,3)$ with reference to the
maximum value of one dimensional case. Moreover, we study the scaling behavior
and the entanglement exponent. We compare the results for one, two and three dimensional
cases and illustrate how the system evolves near the critical point.
\end{titlepage}

\section{INTRODUCTION}

Spin systems have the central importance, regarding the study of
entanglement from the perspectives of quantum information
theory (QIT) and condense matter physics. Spin as a quantum bit is a
miraculous entity which has a pivotal role in the realization of quantum
computers \cite{nc, science1,rmp1, rmp2}. But the main problem arises when we try
to analyse the collective behavior of the infinitely large systems. Quantum
renormalization group (QRG) method is an effective technique in order to
address such problems, i.e. analytically in one dimension \cite{ising, xyz, xxz,
xxzdm, xxznn, xy, xydm, fisher, xxzglobal} and numerically in higher dimensions. 
In the past, different numerical techniques were used to study such systems \cite%
{vedral1,vedral2}. Because numerical techniques are suitable to handle the
computational complexity in finding the ground states, renormalized control
parameters, entanglement and critical properties in higher dimensions \cite%
{vidal1, vidal2, 2dising}.

Many useful methods has been established for the study of strongly correlated 
quantum systems \cite{Amico-2008, wilson1, dmrgrmp, njp1}. Among these techniques,
the density-matrix renormalization group method (DMRG) \cite{white1, white2, white3}
is a reliable and precise numerical technique for the analysis and understanding of 
the low energy properties of such systems in the real space \cite{dmrgrmp}. The DMRG 
method has been used to study the entanglement and the quantum phase transition from 
quantum information entropy in Heisenberg spin systems \cite{dmrgqit}. 
The modified DMRG scheme in a quantum system is shown to preserve the entanglement 
as compare to other numerical techniques \cite{osborne1}. Several computational methods 
were used to study ground states and finite temperature properties of the 
spin systems \cite{sandvik, vidal3}.

Kadanoff's block approach exquisitely participates in the QRG scheme by
dividing the entire lattice into independent blocks whose Hamiltonian is
diagnalized to obtain the lowest energy degenerate states which are further
used for the construction of the renormalized Hilbert space in the lower dimensions \cite%
{kadanoff}. Recently, we extended the application of Kadanoffs block
approach from the one dimension to the two dimensional Hisenberg $XY$ model \cite%
{xy2d} . From the previous study\cite{xy2d}, we realize that the symetrical
extension to the three-dimensional spin system is quite possible and is
presented here. Where we developed the two-dimensional five-spins Kadanoff
block from the one-dimensional three-spins kadanoff block. In this study the
progression from the two-dimensional five-spins Kadanoff block to
the three-dimensional seven-spins Kadanoff block is presented.

The study of quantum correlations in the spin systems through the QRG reflects
both quantum information properties as well as critical properties of the
system \cite{xy2d, ising, xy, xyz, vidal3, osterloh, isingxy}. The ground-state spin
entanglement in d-dimensional bipartite lattice of the $XXZ$ moel is studied \cite{xxzddim},
where bipartite concurrence shows similar qualitative behavior. It reaches to
maximum value at the critical point and it can be seen that this maximum value is
smaller in three dimensions as compared to its two-dimensional counterpart.
Likewise scenario can be seen for the qualitative and the quantitative behavior of the concurrence
in the $XY$ model, as we go from the lower to higher dimensions \cite{xy, xy2d}.
It is the monogamy that limits the entanglement shared among the number of neighbor
sites \cite{Amico-2008}. Similarly, the symmetry of the Hamiltonian predicts that the
magnitude of the entanglement may further reduce in three dimensions due to
increase in the interactions. Moreover, the monogamy is responsible for
achieving the critical point more rapidly, as less QRG iterations are
required in higher dimensions \cite{xy2d}. This critical phenomenon is
described by the behavior of the ground state entanglement as the system size
increases.

In the study presented here, a numerical approach is adopted for the RG evolution of the spin
-1/2 $XY$ model for all the spatial dimensions. Symmetries of the system allow
us to span the spin lattices via Kadanoff's block approach by considering
three, five and seven spins in one, two and three-dimensions respectively
(Fig. 1.), which is necessary in the context of obtaining degenerate ground states 
for the $XY$ model. In this way, the entire lattice is comprised of blocks, where
the each block interacts with all its nearest neighbors through the interblock
interactions (Fig. 2.). We compute the lowest energy eigenvalues of each
Kadanoff block and construct the density matrix from the ground states of
the system. We choose the geometric average of the bipartite concurrence as
the entanglement measure because each interblock spin-spin interaction can be 
seen as bipartite interaction (Fig. 2.). Here, it is worth
mentioning that the reduced density matrices for the different bipartite
interblock spin-spin interactions in two and three dimensions are exactly same which
is the manifestation of the symmetries possessed by the model in all dimensions.
Quantitatively, it is found that for a given spatial dimension, the result
obtained from the geometric average of the bipartite concurrences of
possible interblock spin-spin interactions is equal to the any of the single bipartite
concurrence. The evolution of the entanglement displays the comparative
behavior in this study. Finally, we discuss the scaling behavior by
observing the absolute maximum value of the derivative of the concurrence
and by measuring the entanglement exponent which portray that how the
system approaches critical point in all dimensions.

\begin{figure}
\includegraphics[scale=0.75]{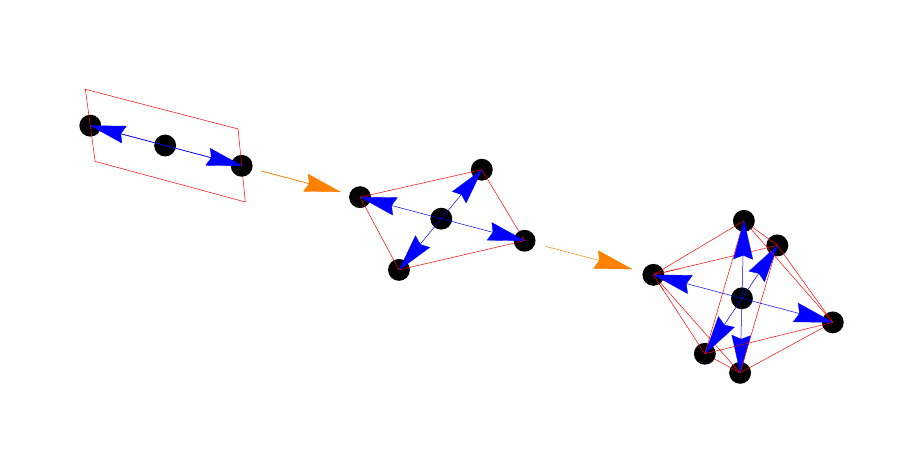}
\caption{\sf (Color online) Symmetric
transformation of Kadanoff's block for one (linear), two (square) and
three dimensional (cubical) spin lattices containing three, five and seven
spins respectively.}
\label{Fig.1.}
\end{figure}

\begin{figure}
\includegraphics[scale=0.75]{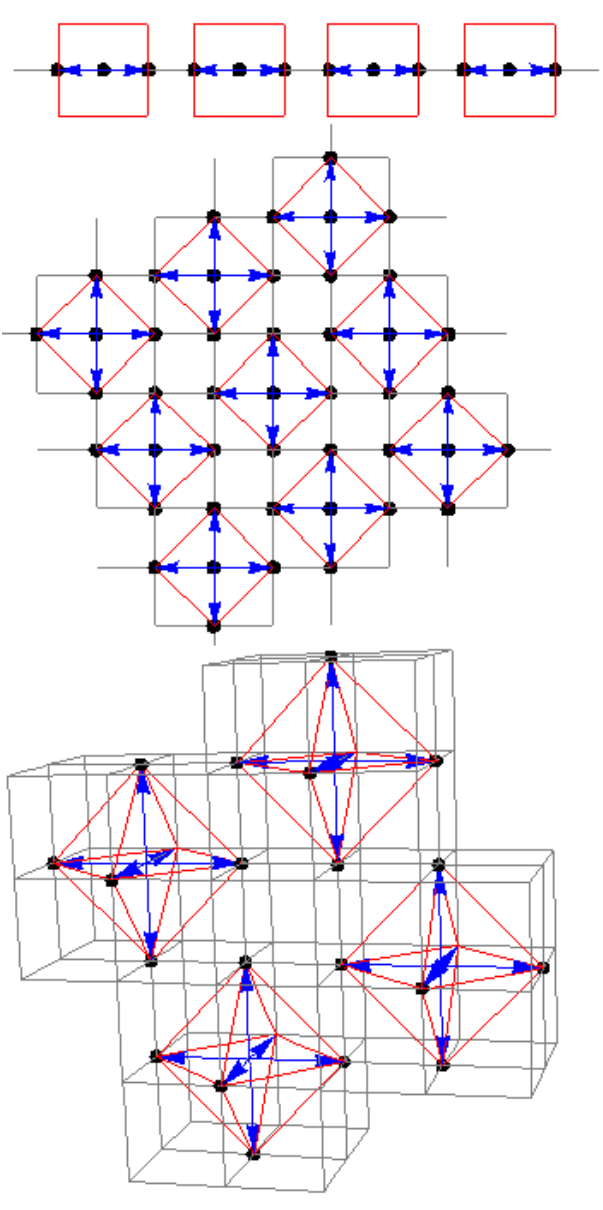}
\caption{\sf (Color online) View of one, two and three dimensional spin lattices spanned by the blocks each
containing three, five and seven spins respectively.
Central spin of each block interacts with all the
remaining spins located at the corner of that block, representing block
interactions whereas neighboring blocks interact via their corner spins,
representing inter-block interactions.}
\label{Fig.2.}
\end{figure}

\section{QRG IMPLEMENTATION}

The QRG technique is a method in which lattice size is reduced in each step
of iteration and reaches a situation where control parameters sustain the
previous renormalized value. Thus the single block with renormalized
coupling constants represents an infinitely large system. The construction
of this block is obtained by using well defined Kadanoff's block approach \cite%
{xy26, xy27}. We consider blocks of three, five and seven spins in one, two
and three dimensions respectively (Fig. 1.) to obtain the effective
Hamiltonian which has similar structure as that of the original Hamiltonian. From the previous 
studies of the $XY$ model \cite{xy, xydm, xy2d}, it is found that 
in the renormalization process, the projection operator constructed from the 
degenerate ground states of the block works well for obtaining the effective Hamiltonian 
in the renormalized Hilbert space of spin -1/2 particle. The degenerate ground eigenstates can 
only be obtained if we consider the blocks containing odd number of spins in any spatial dimensions.
In the construction of Kadanoff's block we make symmetrical transformation from the lower to higher
dimensions with respect to the central spin of the block, i.e., in two dimensions 
two nearest neighbors are added in $y$-direction while in three dimensions further two nearest
neighbors are added in the the $z$-direction to the central spin of
the one-dimensional Kaddanoff's block (Fig. 1.).

The generalized form of the Hamiltonian in $d$-dimensions, for $%
\prod\limits_{d}N_{d},$ spins where $N_{d}=N$ with $d=1,2,3$ for one, two
and three dimensions respectively, can be represented by,

\begin{eqnarray}
H_{d} &=&\frac{J}{4}\prod\limits_{p=1}^{d}S_{p}[\sum_{q=x,y}(1+\gamma
_{q})(c_{1}\sigma _{c_{1}i,c_{2}j,c_{3}k}^{q}\sigma
_{c_{1}(i+1),c_{2}j,c_{3}k}^{q}  \nonumber \\
&&+c_{2}\sigma _{c_{1}i,c_{2}j,c_{3}k}^{q}\sigma
_{c_{1}i,c_{2}(j+1),c_{3}k}^{q}+c_{3}\sigma
_{c_{1}i,c_{2}j,c_{3}k}^{q}\sigma _{c_{1}i,c_{2}j,c_{3}(k+1)}^{q})],
\label{1}
\end{eqnarray}

Where, $\gamma _{x}=\gamma ,\gamma _{y}=-\gamma ,S_{p}=\sum_{r=1}^{N_{d}}$
with $r=i,j$ and $k$ for $p=1,2$ and $3$ respectively along with the
constants $c_{1}=1,c_{2}=(d-1)/2^{((d-1)(d-2)/2)}$ and $c_{3}=(d-1)(d-2)/2$. 
$\sigma ^{x}$ and $\sigma ^{y}$ are the Pauli spin matrices where as $J$ and 
$\gamma $ represent the coupling strength and anisotropic parameter
respectively. The sign of the coupling strength $J$ determines whether the
model lies in antiferromagnetic or ferromagnetic regime. We use 
the positive value of $J$ which corresponds to antiferromagnetic case. However, the sign of $j$
does not effect the results. The quantum fluctuations are driven by the values 
of control parameter $\gamma $. Different values of $\gamma$ reduce the model to different 
classes. i.e., at $\gamma =1$ the system is Ising, for $\gamma =0$ it is $XX$ and for all other
values it is Ising universal class.

For renormalization process, we split the original Hamiltonian $H_{d}$ in to
two parts, the block $H_{d}^{B}$ and the interblock $H_{d}^{BB}$ \cite{ising,
xxz, xxzdm, xy, xydm, xy2d}. The effective Hamiltonian $H_{d}^{eff}$ is
obtained from the $\pi ^{\dagger }H_{d}\pi $, where $\pi =\left\vert \Uparrow
\right\rangle \left\langle \phi _{0}^{1}\right\vert +\left\vert \Downarrow
\right\rangle \left\langle \phi _{0}^{2}\right\vert $ is the projection
operator obtained by projecting the $2^{n}$ (where $n=3,5,7$ for one , two
and three dimensions respectively) dimensional degenerate ground states $%
\left\vert \phi _{0}^{1}\right\rangle $ and $\left\vert \phi
_{0}^{2}\right\rangle $ of the block on spin-$1/2$ qubits $\left\vert
\Uparrow \right\rangle $ and $\left\vert \Downarrow \right\rangle $
resulting in a projection operator in $2$-dimensional Hilbert space.
Finally, we obtained the renormalized numerical values of the coupling
constant $j^{\prime}$ and the anisotropy parameter $\gamma^{\prime}$ from the 
effective Hamiltonian. Some numerical values of $\gamma$ for different RG steps in
the real-space are given in TABLE.1. The solution for $\gamma =\gamma^{\prime}$ 
can be found by plotting $\gamma^{\prime}$ against $\gamma $ for the first RG step
iteration (Fig.3.), which shows that the model attains two different phases 
for $\gamma \rightarrow -1$ or $+1$ referring to Ising like phase 
and $\gamma \rightarrow 0,$ to spin fluid phase suggesting that there lies 
a phase boundary between these two phases. We can see from the plot (Fig.3.) 
that the phase boundary is more prominent in higher dimensions even in the 
first RG iteration compared to the corresponding system in lower dimensions. 
Hence the model demands less number of RG iterations in higher dimensions to reach at the critical point.

\begin{table*}
\caption{\label{tab:table1} A few values of the anisotropic parameter $\gamma$ for $0^{th}$, $1^{st}$ and $2^{nd}$
order RG iterations in $1D$, $2D$ and $3D$ spin-$1/2$ $XY$ model.}
\begin{ruledtabular}
\begin{tabular}{cccccccccc}
&\multicolumn{3}{c}{$0^{th}$ Step RG iteration}&\multicolumn{3}{c}{$1^{st}$ Step RG iteration}&\multicolumn{3}{c}{$2^{nd}$ Step RG iteration}\\
Initial $\gamma$&$1D$&$2D$&$3D$ &$1D$&$2D$&$3D$&$1D$&$2D$&$3D$\\ \hline
-1.0&-1.0&-1.0&-1.0&-1.0&-1.0&-1.0&-1.0&-1.0&-1.0 \\
-0.96&-0.96&-0.96&-0.96&-0.999983&-1.0&-1.0&-1.0&-1.0&-1.0 \\
-0.76&-0.76&-0.76&-0.76&-0.994941&-1.0&-1.0&-1.0&-1.0&-1.0 \\
-0.51&-0.51&-0.51&-0.51&-0.933916&-0.999898&-1.0&-0.99992&-1.0&-1.0 \\
-0.26&-0.26&-0.26&-0.26&-0.663099&-0.989406&-0.999821&-0.983511&-1.0&-1.0 \\
-0.01&-0.01&-0.01&-0.01&-0.029992&-0.109531&-0.225734&-0.0897608&-0.825471&-0.999508 \\
0.04&0.04&0.04&0.04&0.11949&0.41218&0.717215&0.345383&0.999333&1.0 \\
0.24&0.24&0.24&0.24&0.625703&0.984742&0.999678&0.975885&1.0&1.0 \\
0.49&0.49&0.49&0.49&0.922891&0.999848&1.0&0.999871&1.0&1.0 \\
0.74&0.74&0.74&0.74&0.993349&1.0&1.0&1.0&1.0&1.0 \\
0.95&0.95&0.95&0.95&0.999966&1.0&1.0&1.0&1.0&1.0 \\
1.0&1.0&1.0&1.0&1.0&1.0&1.0&1.0&1.0&1.0 \\
			
\end{tabular}
\end{ruledtabular}
\end{table*}

\begin{figure}
\includegraphics[scale=0.75]{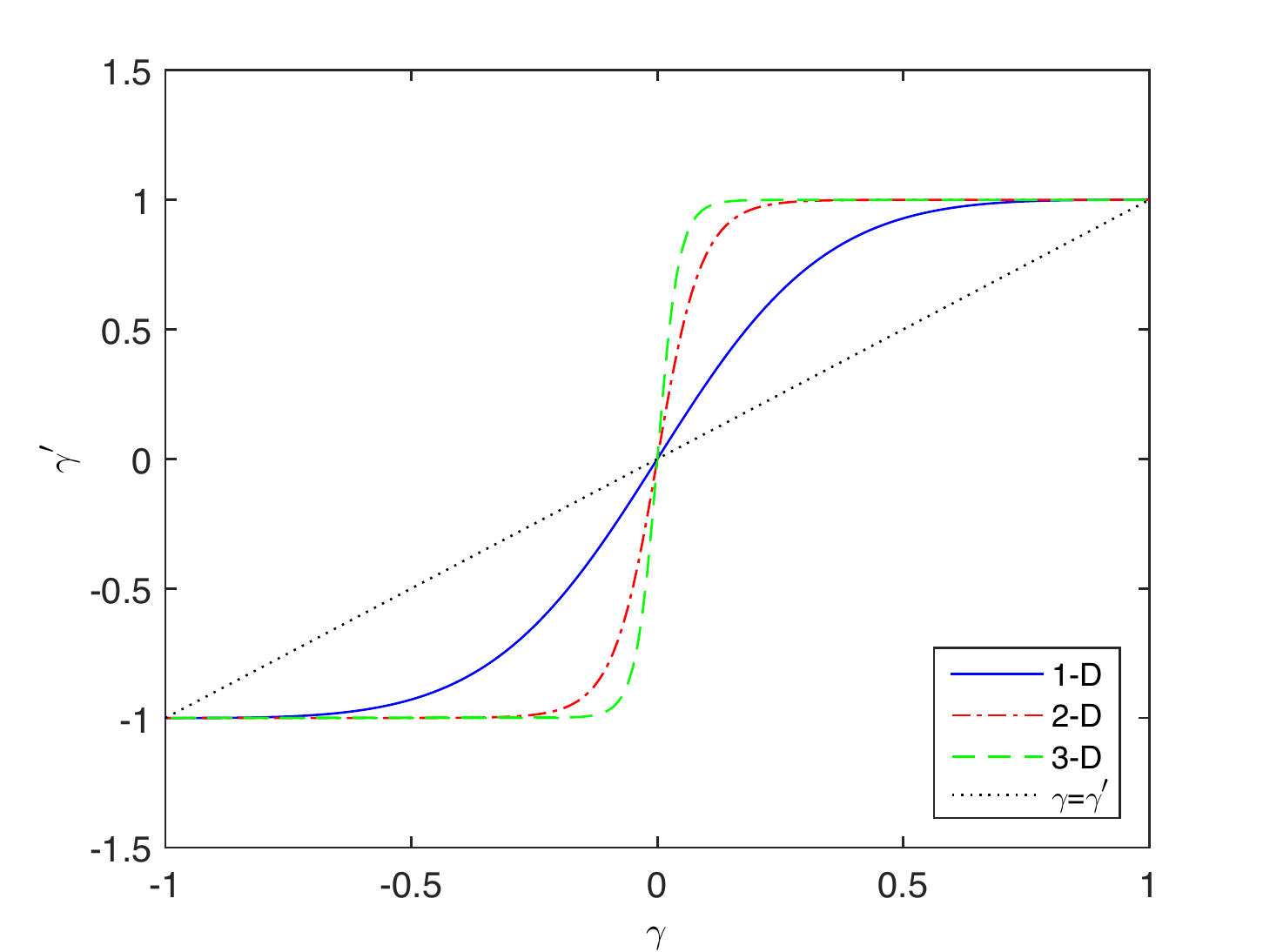}
\caption{\sf (Color online) Comparative plot of $\gamma^{\prime}$ against the anisotropic parameter $\gamma$ for the $1^{st}$ order RG	iteration in one, two and three dimensions, showing the phase variations between two phases corresponding to the values $\gamma =0$ and $\pm 1$ which are obtained from the solution for $\gamma =\gamma^{\prime}$.}
\label{Fig.3.}
\end{figure}

\section{COMPARATIVE STUDY}

To perform the comparative study of the $XY$-model in the real space by the
QRG technique, we choose the concurrence as the entanglement measure \cite%
{wootters}. In one dimension, bipartite interblock spin-spin
interaction exists through out the spin chain, while in higher dimensions
many bipartite interblock interactions are present. From this perspective it
may be appropriate to use the geometric average of the bipartite
concurrences as a generalized criterion for the entanglement study. Further
we find that the value of each bipartite concurrence for any particular
iteration is same and equal to the geometric average of the bipartite
concurrences. Therefore, keeping in view this scenario, we can use a single
bipartite concurrence as the entanglement measure. The bipartite concurrence 
$C_{ij}$ for $i^{th}$ and $j^{th}$ nearest neighbor spins obtained 
from reduced density matrix $\rho _{ij}$ is given as

\begin{equation}
C_{ij}=\max [\sqrt{\lambda _{ij,4}}-\sqrt{\lambda _{ij,3}}-\sqrt{\lambda
_{ij,2}}-\sqrt{\lambda _{ij,1}},0],  \label{2}
\end{equation}

where $\lambda _{ij,k}$ for $(k=1,2,3,4)$ are the eigenvalues of matrix $%
\rho _{ij}\tilde{\rho}_{ij}$ with $\tilde{\rho}_{ij}=(\sigma _{i}^{y}\otimes
\sigma _{j}^{y})$\ $\rho _{ij}^{\ast }(\sigma _{i}^{y}\otimes \sigma
_{j}^{y})$ and $\lambda _{ij,4}>\lambda _{ij,3}>\lambda _{ij,2}>\lambda $.
The reduced density matrix $\rho _{ij}$ for the spins interaction $i$ and $j$
is constructed by taking the marginal partial traces of the total density
matrix $\rho $ where, $\rho =\left\vert \phi _{0}^{1}\right\rangle
\left\langle \phi _{0}^{1}\right\vert $ is obtained from any of the
degenerate ground state of the Hamiltonian.

The numerical results of the concurrence $C_{ij}$ are plotted against the
anisotropy parameter $\gamma $ (Fig. 4.) for the first three RG iterations
of the spin-$1/2$ $XY$ model in all dimensions. The concurrence shows
similar qualitative behavior but smaller in magnitude in higher dimensions 
as compare to lower dimensions. At $\gamma =0$, the concurrence attains maximum 
values which are $0.5$, $0.25$ and $0.167$ in one, two and three dimensions respectively.
We find that the maximum value of concurrence decreases by the factor of $1/n$ $%
(n=2,3)$ compared to the one-dimensional case. It can be interpreted that the
concurrence decreases with the increase in monogamy in the higher dimensions. In
all dimensions, in the final iteration the system reaches thermodynamic
limit and the concurrence acquires two fixed values, one non zero at the
critical point ($\gamma =0)$ favoring long range disordered phase and the
other at $\gamma \neq 0$, where the concurrence is zero and the system
corresponds to a dominant ordered phase. The phenomenon of phase transition
for the $XY$ model in the real space can also be seen through the non
analytic behavior of the concurrence where the derivative of the concurrence 
shows the discontinuity at the critical point (Fig. 4.). Whereas the
concurrence itself remains continuous as the system size becomes infinitely large.
Such phase transitions are named as the second-order QPT.

\begin{figure}
\includegraphics[scale=0.75]{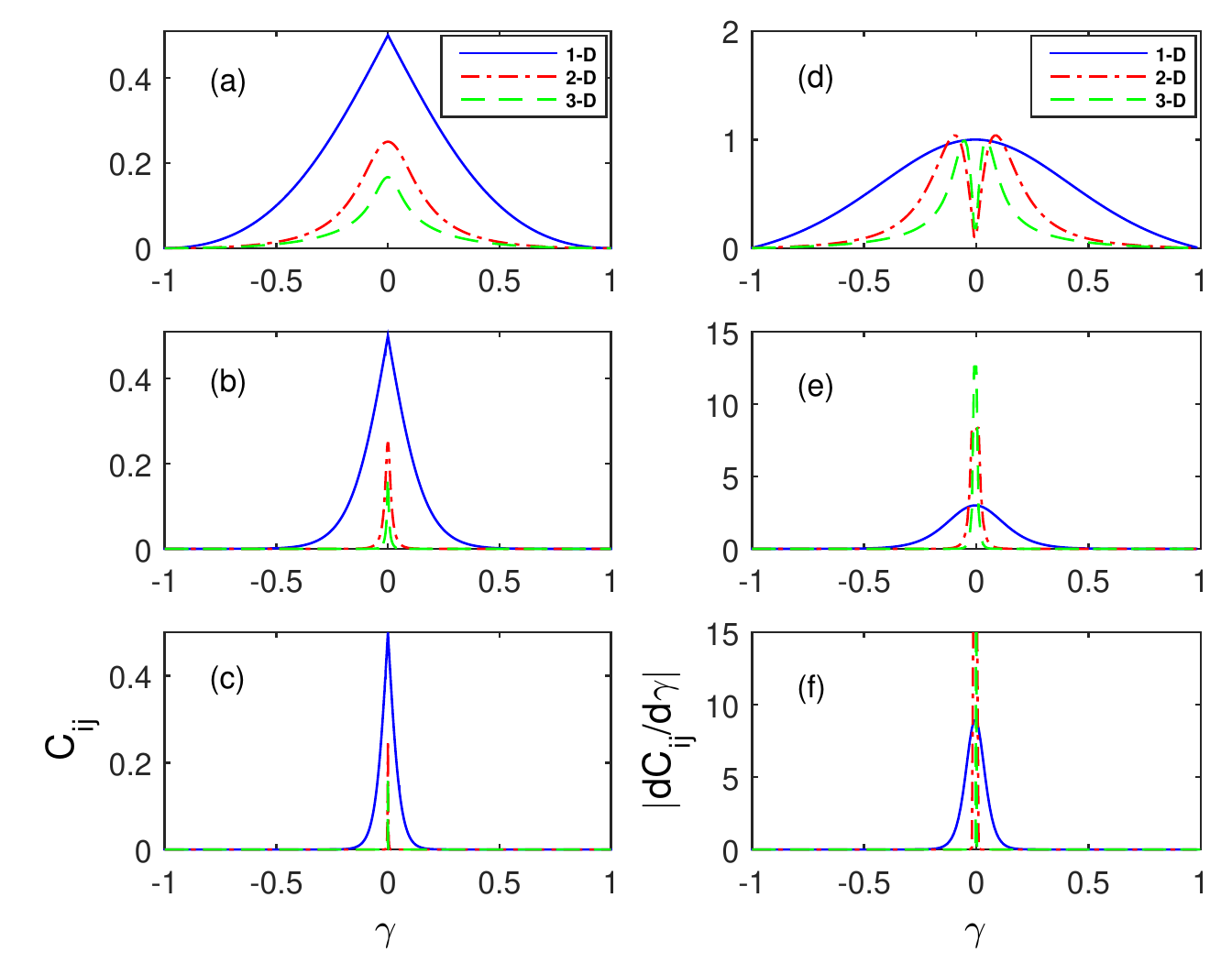}
\caption{\sf (Color online) Evolution of
the concurrence $C_{ij}$ (where $i$ and $j$ represents the nearest neighbor spins) figures (a),(b) and (c), and its absolute derivative $%
\left\vert dC_{ij}/d\protect\gamma \right\vert $ figures (d), (e) and (f), against the
anisotropic parameter $\protect\gamma $ for the zeroth (a) and (d), first (b) and (e) and second (c) and (f) RG
steps for the one (solid line), two (dash-dot line) and three-dimensional (dashed line) systems are shown
here. The concurrence shows the same qualitative behavior but its peak value
reduces with the dimensionality of the system, while the derivative of the
concurrence diverges more rapidly.}
\label{Fig.4.}
\end{figure}

The detail analysis of the entanglement as a resource for exploring 
the critical phenomenon in the one-dimensional $XY$ model has been done in Ref. \cite{osterloh, osborne2}.
It was cojectured that the system is maximally entangled at the critical point
corresponding to the delocalized state, where the correlations exist on all length
scales, in contrast to the situation away from the critical point favoring
the exponential decay of the correlations as a function of sites separation \cite{sachdev}.
Our results depict similar behavior for the entanglement in the vicinity of the critical point even in
all the spatial dimensions. These numerical results of entanglement are also supported by the previous
analytical results of $1D$ and $2D$ $XY$ model \cite{xy, xy2d}.
In Ref. \cite{osterloh,osborne2}, the transverse Ising model was considered 
as a special case of the $XY$ model for anisotropy parameter $\gamma=1$. In the 
limiting case of the transverse Ising model, where the effect of external 
field is very small the value of entanglement approaches zero conforming
to our result of Ising phase for $\gamma=\pm1$. The degenerate ground states at $\gamma=1$ 
for $XY$ model for the second step RG iteration are given in the appendix.

The emergence of singularity by the divergence of the derivative can be
probed by investigating the scaling behavior which relates maximum/minimum
of the derivative of entanglement with the system size. From the comparative
plots (Fig. 5.), it is observed that in each dimension, the derivative of the
concurrence shows linear behavior on the logarithmic scale but the slope
increases with increasing the dimensionality of the system. Which indicates
that the system diverges rapidly in higher dimensions. To analyze that how
fast the system reaches the transition point, the $\ln (\gamma _{c}-\gamma
_{m})$ is plotted against the $\ln (N)$ (Fig.6.). The value of the
entanglement exponent $\theta $ can be found from the relation $\gamma
_{m}=\gamma _{c}-N^{-\theta }$, which are are $0.73$, $1.48$ and $1.60$ for 
one, two and three dimensions respectively. It can be seen that the entanglement exponent is
highest in three dimensions, which corresponds to smaller correlation length, supporting the idea that it
reaches the thermodynamic limit rapidly and acquires phase transition in
less number of RG iterations. It is worth mentioning that the divergence of the derivative of the concurrence in the vicinity of the quantum phase transition point was investigated in one-dimensional $XY$ model with different lattice sizes in Ref. \cite{osterloh} and with QRG technique in Ref. \cite{xy, xydm, xy2d}. Complying to these studies, the entanglement in our study, also obeys the scaling behavior near the critical point.

\begin{figure}
\includegraphics[scale=0.75]{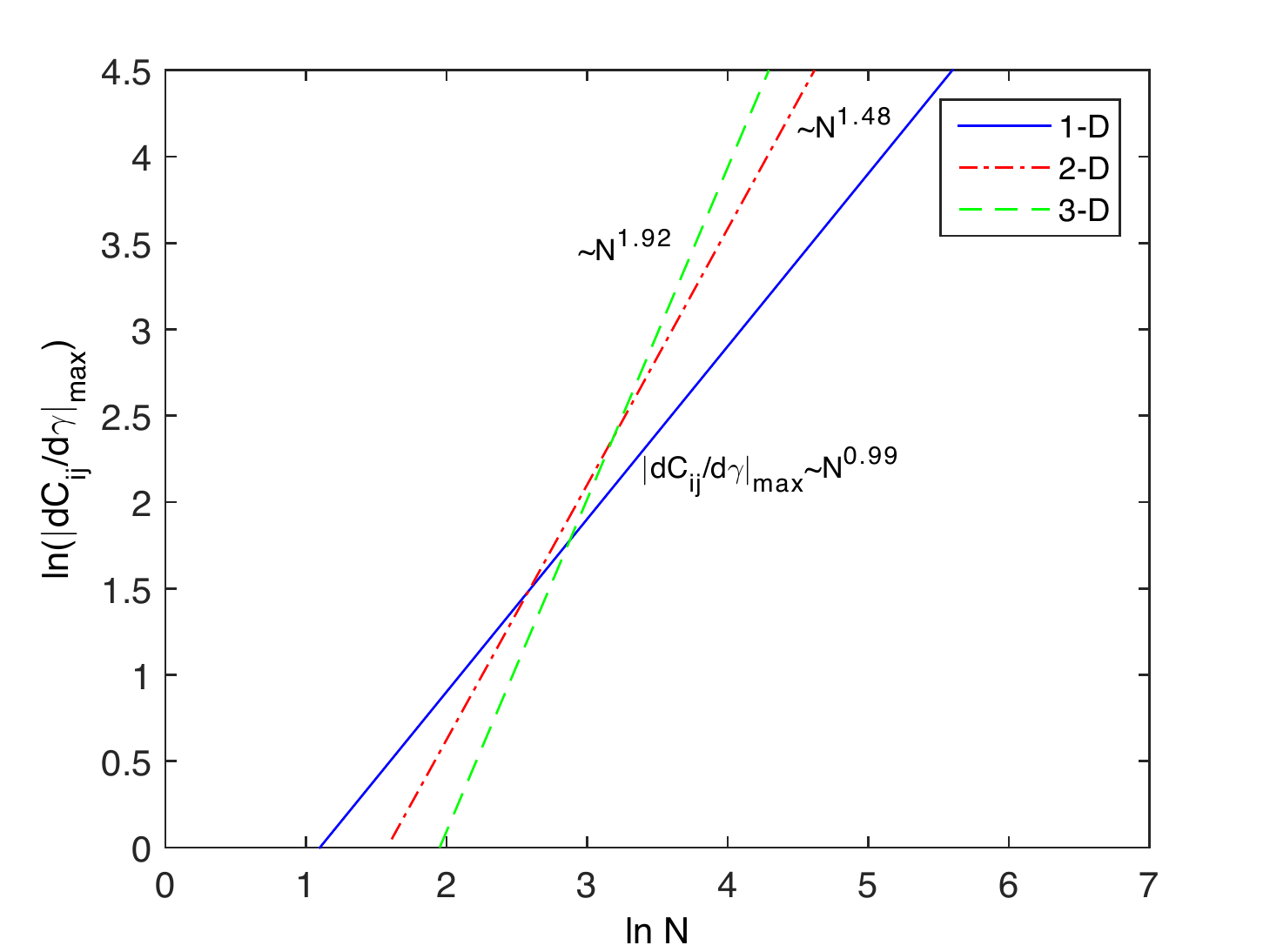}
\caption{\sf (Color online) The scaling
behavior of the concurrence is depicted by plotting $\ln (\left\vert dC_{ij}/d%
\protect\gamma \right\vert _{\max })$ against $\ln N$. It shows linear
behavior where slope increases with the dimensionality of the model.}
\label{Fig.5.}
\end{figure}

\begin{figure}
\includegraphics[scale=0.75]{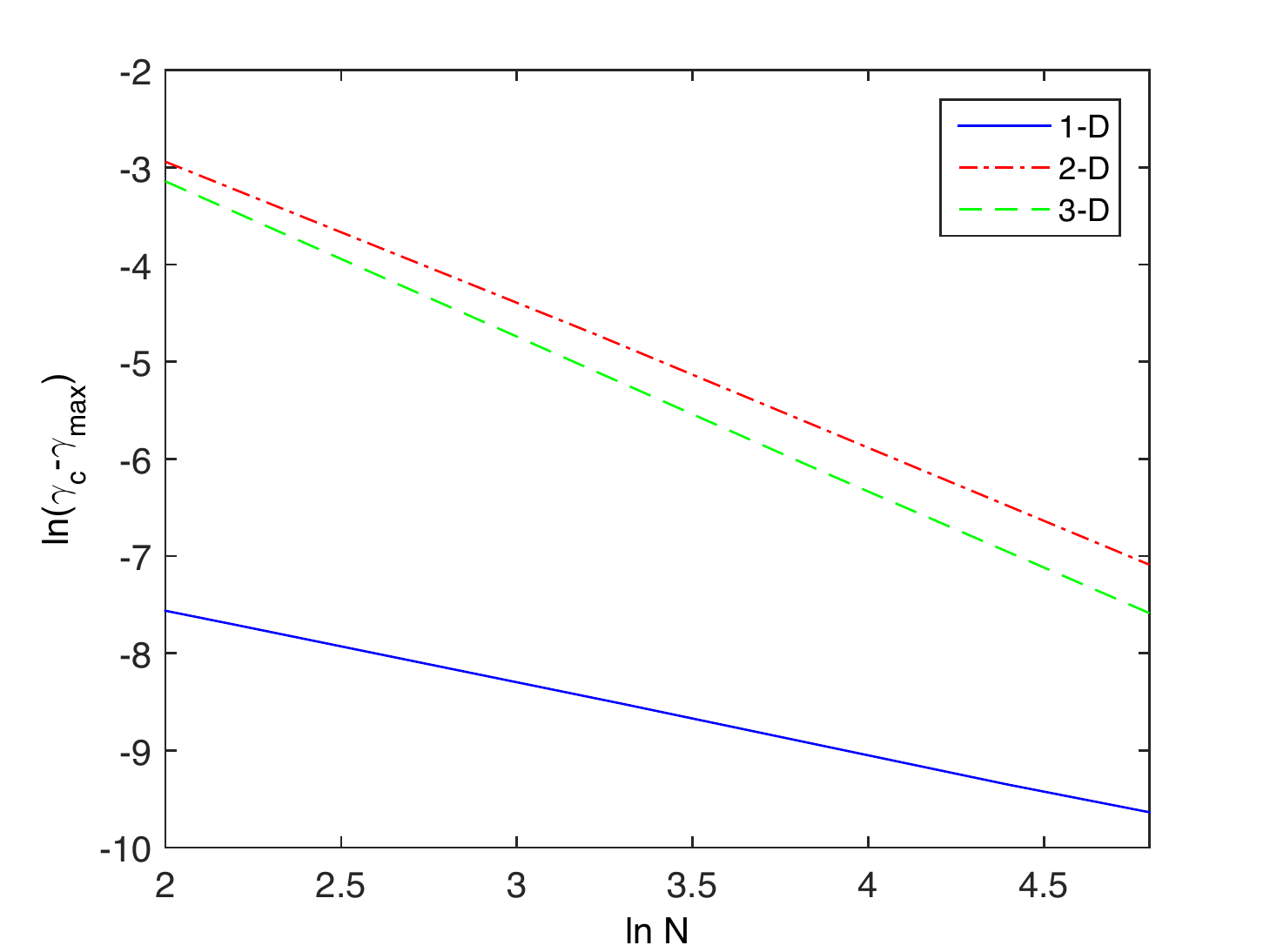}
\caption{\sf (Color online) Behavior of the $%
\protect\gamma _{c}-\protect\gamma _{\max }$ against the system size $N$
showing rapid approach of the $\protect\gamma _{\max }$ to the critical point as the
system dimensionality increases.}
\label{Fig.6.}
\end{figure}

\section{CONCLUSIONS}

The QRG in the real-space for the spin -$1/2$ $XY$ model is achieved. We see that
for the concurrence as the entanglement measure, the system shows similar
qualitative behavior in all dimensions, which is the manifestation of the
symmetric interactions present in the model. But the maximum value of the
entanglement in two and three dimensional $XY$ model is decreased by the
factor of $1/n$, where $n=2, 3$, with reference to the one-dimensional case.
The reduction in the peak value is due to the increase in the monogamy of the spins
interactions. Our results are consistent  with the previous studies of the one and two-dimensional
$XY$ model.  This study will be helpful in understanding the behavior of the entanglement and the critical phenomenon for different types of complex spin systems in real-space.

\section{ACKNOWLEDGMENTS}

This work is partly supported by the HIGHER EDUCATION COMMISSION, PAKISTAN
under the Indigenous Ph.D. Fellowship Scheme.
\section{APPENDIX}

The expression for degenerate ground state eigenvectors $|\phi(1)\rangle$ and $|\phi(2)\rangle$ of
the $1D$, $2D$ and  $3D$ $XY$ model at $\gamma=1$ corresponding to the Ising class are given below;

\begin{eqnarray}
|\phi_{1}\rangle^{1D}=-0.5|\uparrow\uparrow\uparrow\rangle+0.5|\uparrow\downarrow\downarrow\rangle-0.5|\downarrow\uparrow\downarrow\rangle+0.5|\downarrow\downarrow\uparrow\rangle\nonumber
\end{eqnarray}	
\begin{eqnarray}
|\phi_{2}\rangle^{1D}=-0.5|\uparrow\uparrow\downarrow\rangle+0.5|\uparrow\downarrow\uparrow\rangle-0.5|\downarrow\uparrow\uparrow\rangle+0.5|\downarrow\downarrow\downarrow\rangle\nonumber
\end{eqnarray}
\begin{eqnarray}
|\phi_{1}\rangle^{2D}=-0.25|\uparrow\uparrow\uparrow\uparrow\downarrow\rangle-0.25|\uparrow\uparrow\uparrow\downarrow\uparrow\rangle-0.25|\uparrow\uparrow\downarrow\uparrow\uparrow\rangle-0.25|\uparrow\uparrow\downarrow\downarrow\downarrow\rangle-0.25|\uparrow\downarrow\uparrow\uparrow\uparrow\rangle-0.25|\uparrow\downarrow\uparrow\downarrow\downarrow\rangle-0.25|\uparrow\downarrow\downarrow\uparrow\downarrow\rangle-\nonumber
\\0.25|\uparrow\downarrow\downarrow\downarrow\uparrow\rangle+0.25|\downarrow\uparrow\uparrow\uparrow\uparrow\rangle+0.25|\downarrow\uparrow\uparrow\downarrow\downarrow\rangle+0.25|\downarrow\uparrow\downarrow\uparrow\downarrow\rangle+0.25|\downarrow\uparrow\downarrow\downarrow\uparrow\rangle+0.25|\downarrow\downarrow\uparrow\uparrow\downarrow\rangle+0.25|\downarrow\downarrow\uparrow\downarrow\uparrow\rangle+\nonumber
\\0.25|\downarrow\downarrow\downarrow\uparrow\uparrow\rangle+0.25|\downarrow\downarrow\downarrow\downarrow\downarrow\rangle\nonumber
\end{eqnarray}
\begin{eqnarray}
|\phi_{2}\rangle^{2D}=0.25|\uparrow\uparrow\uparrow\uparrow\uparrow\rangle+0.25|\uparrow\uparrow\uparrow\downarrow\downarrow\rangle+0.25|\uparrow\uparrow\downarrow\uparrow\downarrow\rangle+0.25|\uparrow\uparrow\downarrow\downarrow\uparrow\rangle+0.25|\uparrow\downarrow\uparrow\uparrow\downarrow\rangle+0.25|\uparrow\downarrow\uparrow\downarrow\uparrow\rangle+0.25|\uparrow\downarrow\downarrow\uparrow\uparrow\rangle+\nonumber
\\0.25|\uparrow\downarrow\downarrow\downarrow\downarrow\rangle-0.25|\downarrow\uparrow\uparrow\uparrow\downarrow\rangle-0.25|\downarrow\uparrow\uparrow\downarrow\uparrow\rangle-0.25|\downarrow\uparrow\downarrow\uparrow\uparrow\rangle-0.25|\downarrow\uparrow\downarrow\downarrow\downarrow\rangle-0.25|\downarrow\downarrow\uparrow\uparrow\uparrow\rangle-0.25|\downarrow\downarrow\uparrow\downarrow\downarrow\rangle-\nonumber
\\0.25|\downarrow\downarrow\downarrow\uparrow\downarrow\rangle-0.25|\downarrow\downarrow\downarrow\downarrow\uparrow\rangle\nonumber
\end{eqnarray}
\begin{eqnarray}
|\phi_{1}\rangle^{3D}=0.125|\uparrow\uparrow\uparrow\uparrow\uparrow\uparrow\uparrow\rangle+
0.125|\uparrow\uparrow\uparrow\uparrow\uparrow\downarrow\downarrow\rangle+
0.125|\uparrow\uparrow\uparrow\uparrow\downarrow\uparrow\downarrow\rangle+
0.125|\uparrow\uparrow\uparrow\uparrow\downarrow\downarrow\uparrow\rangle+
0.125|\uparrow\uparrow\uparrow\downarrow\uparrow\uparrow\downarrow\rangle+
0.125|\uparrow\uparrow\uparrow\downarrow\uparrow\downarrow\uparrow\rangle+\nonumber
\\0.125|\uparrow\uparrow\uparrow\downarrow\downarrow\uparrow\uparrow\rangle+
0.125|\uparrow\uparrow\uparrow\downarrow\downarrow\downarrow\downarrow\rangle+
0.125|\uparrow\uparrow\downarrow\uparrow\uparrow\uparrow\downarrow\rangle+
0.125|\uparrow\uparrow\downarrow\uparrow\uparrow\downarrow\uparrow\rangle+
0.125|\uparrow\uparrow\downarrow\uparrow\downarrow\uparrow\uparrow\rangle+
0.125|\uparrow\uparrow\downarrow\uparrow\downarrow\downarrow\downarrow\rangle+\nonumber
\\0.125|\uparrow\uparrow\downarrow\downarrow\uparrow\uparrow\uparrow\rangle+
0.125|\uparrow\uparrow\downarrow\downarrow\uparrow\downarrow\downarrow\rangle+
0.125|\uparrow\uparrow\downarrow\downarrow\downarrow\uparrow\downarrow\rangle+
0.125|\uparrow\uparrow\downarrow\downarrow\downarrow\downarrow\uparrow\rangle+
0.125|\uparrow\downarrow\uparrow\uparrow\uparrow\uparrow\downarrow\rangle+
0.125|\uparrow\downarrow\uparrow\uparrow\uparrow\downarrow\uparrow\rangle+\nonumber
\\0.125|\uparrow\downarrow\uparrow\uparrow\downarrow\uparrow\uparrow\rangle+
0.125|\uparrow\downarrow\uparrow\uparrow\downarrow\downarrow\downarrow\rangle+
0.125|\uparrow\downarrow\uparrow\downarrow\uparrow\uparrow\uparrow\rangle+
0.125|\uparrow\downarrow\uparrow\downarrow\uparrow\downarrow\downarrow\rangle+
0.125|\uparrow\downarrow\uparrow\downarrow\downarrow\uparrow\downarrow\rangle+
0.125|\uparrow\downarrow\uparrow\downarrow\downarrow\downarrow\uparrow\rangle+\nonumber
\\0.125|\uparrow\downarrow\downarrow\uparrow\uparrow\uparrow\uparrow\rangle+
0.125|\uparrow\downarrow\downarrow\uparrow\uparrow\downarrow\downarrow\rangle+
0.125|\uparrow\downarrow\downarrow\uparrow\downarrow\uparrow\downarrow\rangle+
0.125|\uparrow\downarrow\downarrow\uparrow\downarrow\downarrow\uparrow\rangle+
0.125|\uparrow\downarrow\downarrow\downarrow\uparrow\uparrow\downarrow\rangle+
0.125|\uparrow\downarrow\downarrow\downarrow\uparrow\downarrow\uparrow\rangle+\nonumber
\\0.125|\uparrow\downarrow\downarrow\downarrow\downarrow\uparrow\uparrow\rangle+
0.125|\uparrow\downarrow\downarrow\downarrow\downarrow\downarrow\downarrow\rangle-
0.125|\downarrow\uparrow\uparrow\uparrow\uparrow\uparrow\downarrow\rangle-
0.125|\downarrow\uparrow\uparrow\uparrow\uparrow\downarrow\uparrow\rangle-
0.125|\downarrow\uparrow\uparrow\uparrow\downarrow\uparrow\uparrow\rangle-
0.125|\downarrow\uparrow\uparrow\uparrow\downarrow\downarrow\downarrow\rangle-\nonumber
\\0.125|\downarrow\uparrow\uparrow\downarrow\uparrow\uparrow\uparrow\rangle-
0.125|\downarrow\uparrow\uparrow\downarrow\uparrow\downarrow\downarrow\rangle-
0.125|\downarrow\uparrow\uparrow\downarrow\downarrow\uparrow\downarrow\rangle-
0.125|\downarrow\uparrow\uparrow\downarrow\downarrow\downarrow\uparrow\rangle-
0.125|\downarrow\uparrow\downarrow\uparrow\uparrow\uparrow\uparrow\rangle-
0.125|\downarrow\uparrow\downarrow\uparrow\uparrow\downarrow\downarrow\rangle-\nonumber
\\0.125|\downarrow\uparrow\downarrow\uparrow\downarrow\uparrow\downarrow\rangle-
0.125|\downarrow\uparrow\downarrow\uparrow\downarrow\downarrow\uparrow\rangle-
0.125|\downarrow\uparrow\downarrow\downarrow\uparrow\uparrow\downarrow\rangle-
0.125|\downarrow\uparrow\downarrow\downarrow\uparrow\downarrow\uparrow\rangle-
0.125|\downarrow\uparrow\downarrow\downarrow\downarrow\uparrow\uparrow\rangle-
0.125|\downarrow\uparrow\downarrow\downarrow\downarrow\downarrow\downarrow\rangle-\nonumber
\\0.125|\downarrow\downarrow\uparrow\uparrow\uparrow\uparrow\uparrow\rangle-
0.125|\downarrow\downarrow\uparrow\uparrow\uparrow\downarrow\downarrow\rangle-
0.125|\downarrow\downarrow\uparrow\uparrow\downarrow\uparrow\downarrow\rangle-
0.125|\downarrow\downarrow\uparrow\uparrow\downarrow\downarrow\uparrow\rangle-
0.125|\downarrow\downarrow\uparrow\downarrow\uparrow\uparrow\downarrow\rangle-
0.125|\downarrow\downarrow\uparrow\downarrow\uparrow\downarrow\uparrow\rangle-\nonumber
\\0.125|\downarrow\downarrow\uparrow\downarrow\downarrow\uparrow\uparrow\rangle-
0.125|\downarrow\downarrow\uparrow\downarrow\downarrow\downarrow\downarrow\rangle-
0.125|\downarrow\downarrow\downarrow\uparrow\uparrow\uparrow\downarrow\rangle-
0.125|\downarrow\downarrow\downarrow\uparrow\uparrow\downarrow\uparrow\rangle-
0.125|\downarrow\downarrow\downarrow\uparrow\downarrow\uparrow\uparrow\rangle-
0.125|\downarrow\downarrow\downarrow\uparrow\downarrow\downarrow\downarrow\rangle-\nonumber
\\0.125|\downarrow\downarrow\downarrow\downarrow\uparrow\uparrow\uparrow\rangle-
0.125|\downarrow\downarrow\downarrow\downarrow\uparrow\downarrow\downarrow\rangle-
0.125|\downarrow\downarrow\downarrow\downarrow\downarrow\uparrow\downarrow\rangle-
0.125|\downarrow\downarrow\downarrow\downarrow\downarrow\downarrow\uparrow\rangle\nonumber
\end{eqnarray}

\begin{eqnarray}
|\phi_{2}\rangle^{3D}=-0.125|\uparrow\uparrow\uparrow\uparrow\uparrow\uparrow\downarrow\rangle-
0.125|\uparrow\uparrow\uparrow\uparrow\uparrow\downarrow\uparrow\rangle-
0.125|\uparrow\uparrow\uparrow\uparrow\downarrow\uparrow\uparrow\rangle-
0.125|\uparrow\uparrow\uparrow\uparrow\downarrow\downarrow\downarrow\rangle-
0.125|\uparrow\uparrow\uparrow\downarrow\uparrow\uparrow\uparrow\rangle-
0.125|\uparrow\uparrow\uparrow\downarrow\uparrow\downarrow\downarrow\rangle-\nonumber
\\0.125|\uparrow\uparrow\uparrow\downarrow\downarrow\uparrow\downarrow\rangle-
0.125|\uparrow\uparrow\uparrow\downarrow\downarrow\downarrow\uparrow\rangle-
0.125|\uparrow\uparrow\downarrow\uparrow\uparrow\uparrow\uparrow\rangle-
0.125|\uparrow\uparrow\downarrow\uparrow\uparrow\downarrow\downarrow\rangle-
0.125|\uparrow\uparrow\downarrow\uparrow\downarrow\uparrow\downarrow\rangle-
0.125|\uparrow\uparrow\downarrow\uparrow\downarrow\downarrow\uparrow\rangle-\nonumber
\\0.125|\uparrow\uparrow\downarrow\downarrow\uparrow\uparrow\downarrow\rangle-
0.125|\uparrow\uparrow\downarrow\downarrow\uparrow\downarrow\uparrow\rangle-
0.125|\uparrow\uparrow\downarrow\downarrow\downarrow\uparrow\uparrow\rangle-
0.125|\uparrow\uparrow\downarrow\downarrow\downarrow\downarrow\downarrow\rangle-
0.125|\uparrow\downarrow\uparrow\uparrow\uparrow\uparrow\uparrow\rangle-
0.125|\uparrow\downarrow\uparrow\uparrow\uparrow\downarrow\downarrow\rangle-\nonumber
\\0.125|\uparrow\downarrow\uparrow\uparrow\downarrow\uparrow\downarrow\rangle-
0.125|\uparrow\downarrow\uparrow\uparrow\downarrow\downarrow\uparrow\rangle-
0.125|\uparrow\downarrow\uparrow\downarrow\uparrow\uparrow\downarrow\rangle-
0.125|\uparrow\downarrow\uparrow\downarrow\uparrow\downarrow\uparrow\rangle-
0.125|\uparrow\downarrow\uparrow\downarrow\downarrow\uparrow\uparrow\rangle-
0.125||\uparrow\downarrow\uparrow\downarrow\downarrow\downarrow\downarrow\rangle-\nonumber
\\0.125|\uparrow\downarrow\downarrow\uparrow\uparrow\uparrow\downarrow\rangle-
0.125|\uparrow\downarrow\downarrow\uparrow\uparrow\downarrow\uparrow\rangle-
0.125|\uparrow\downarrow\downarrow\uparrow\downarrow\uparrow\uparrow\rangle-
0.125|\uparrow\downarrow\downarrow\uparrow\downarrow\downarrow\downarrow\rangle-
0.125|\uparrow\downarrow\downarrow\downarrow\uparrow\uparrow\uparrow\rangle-
0.125|\uparrow\downarrow\downarrow\downarrow\uparrow\downarrow\downarrow\rangle-\nonumber
\\0.125|\uparrow\downarrow\downarrow\downarrow\downarrow\uparrow\downarrow\rangle-
0.125|\uparrow\downarrow\downarrow\downarrow\downarrow\downarrow\uparrow\rangle+
0.125|\downarrow\uparrow\uparrow\uparrow\uparrow\uparrow\uparrow\rangle+
0.125|\downarrow\uparrow\uparrow\uparrow\uparrow\downarrow\downarrow\rangle+
0.125|\downarrow\uparrow\uparrow\uparrow\downarrow\uparrow\downarrow\rangle+
0.125|\downarrow\uparrow\uparrow\uparrow\downarrow\downarrow\uparrow\rangle+\nonumber
\\0.125|\downarrow\uparrow\uparrow\downarrow\uparrow\uparrow\downarrow\rangle+
0.125|\downarrow\uparrow\uparrow\downarrow\uparrow\downarrow\uparrow\rangle+
0.125|\downarrow\uparrow\uparrow\downarrow\downarrow\uparrow\uparrow\rangle+
0.125|\downarrow\uparrow\uparrow\downarrow\downarrow\downarrow\downarrow\rangle+
0.125|\downarrow\uparrow\downarrow\uparrow\uparrow\uparrow\downarrow\rangle+
0.125|\downarrow\uparrow\downarrow\uparrow\uparrow\downarrow\uparrow\rangle+\nonumber
\\0.125|\downarrow\uparrow\downarrow\uparrow\downarrow\uparrow\uparrow\rangle+
0.125|\downarrow\uparrow\downarrow\uparrow\downarrow\downarrow\downarrow\rangle+
0.125|\downarrow\uparrow\downarrow\downarrow\uparrow\uparrow\uparrow\rangle+
0.125|\downarrow\uparrow\downarrow\downarrow\uparrow\downarrow\downarrow\rangle+
0.125|\downarrow\uparrow\downarrow\downarrow\downarrow\uparrow\downarrow\rangle+
0.125|\downarrow\uparrow\downarrow\downarrow\downarrow\downarrow\uparrow\rangle+\nonumber
\\0.125|\downarrow\downarrow\uparrow\uparrow\uparrow\uparrow\downarrow\rangle+
0.125|\downarrow\downarrow\uparrow\uparrow\uparrow\downarrow\uparrow\rangle+
0.125|\downarrow\downarrow\uparrow\uparrow\downarrow\uparrow\uparrow\rangle+
0.125|\downarrow\downarrow\uparrow\uparrow\downarrow\downarrow\downarrow\rangle+
0.125|\downarrow\downarrow\uparrow\downarrow\uparrow\uparrow\uparrow\rangle+
0.125|\downarrow\downarrow\uparrow\downarrow\uparrow\downarrow\downarrow\rangle+\nonumber
\\0.125|\downarrow\downarrow\uparrow\downarrow\downarrow\uparrow\downarrow\rangle+
0.125|\downarrow\downarrow\uparrow\downarrow\downarrow\downarrow\uparrow\rangle+
0.125|\downarrow\downarrow\downarrow\uparrow\uparrow\uparrow\uparrow\rangle+
0.125|\downarrow\downarrow\downarrow\uparrow\uparrow\downarrow\downarrow\rangle+
0.125|\downarrow\downarrow\downarrow\uparrow\downarrow\uparrow\downarrow\rangle+
0.125|\downarrow\downarrow\downarrow\uparrow\downarrow\downarrow\uparrow\rangle+\nonumber
\\0.125|\downarrow\downarrow\downarrow\downarrow\uparrow\uparrow\downarrow\rangle+
0.125|\downarrow\downarrow\downarrow\downarrow\uparrow\downarrow\uparrow\rangle+
0.125|\downarrow\downarrow\downarrow\downarrow\downarrow\uparrow\uparrow\rangle+
0.125|\downarrow\downarrow\downarrow\downarrow\downarrow\downarrow\downarrow\rangle\nonumber
\end{eqnarray}
Both the degenerate eigenvectors are normalized and orthogonal i.e., $\langle\phi_{i}|\phi_{j}\rangle=\delta_{ij}$ 
where $\delta_{ij}$ is the Kronecker delta function.

\begin{figure}
\includegraphics[scale=0.75]{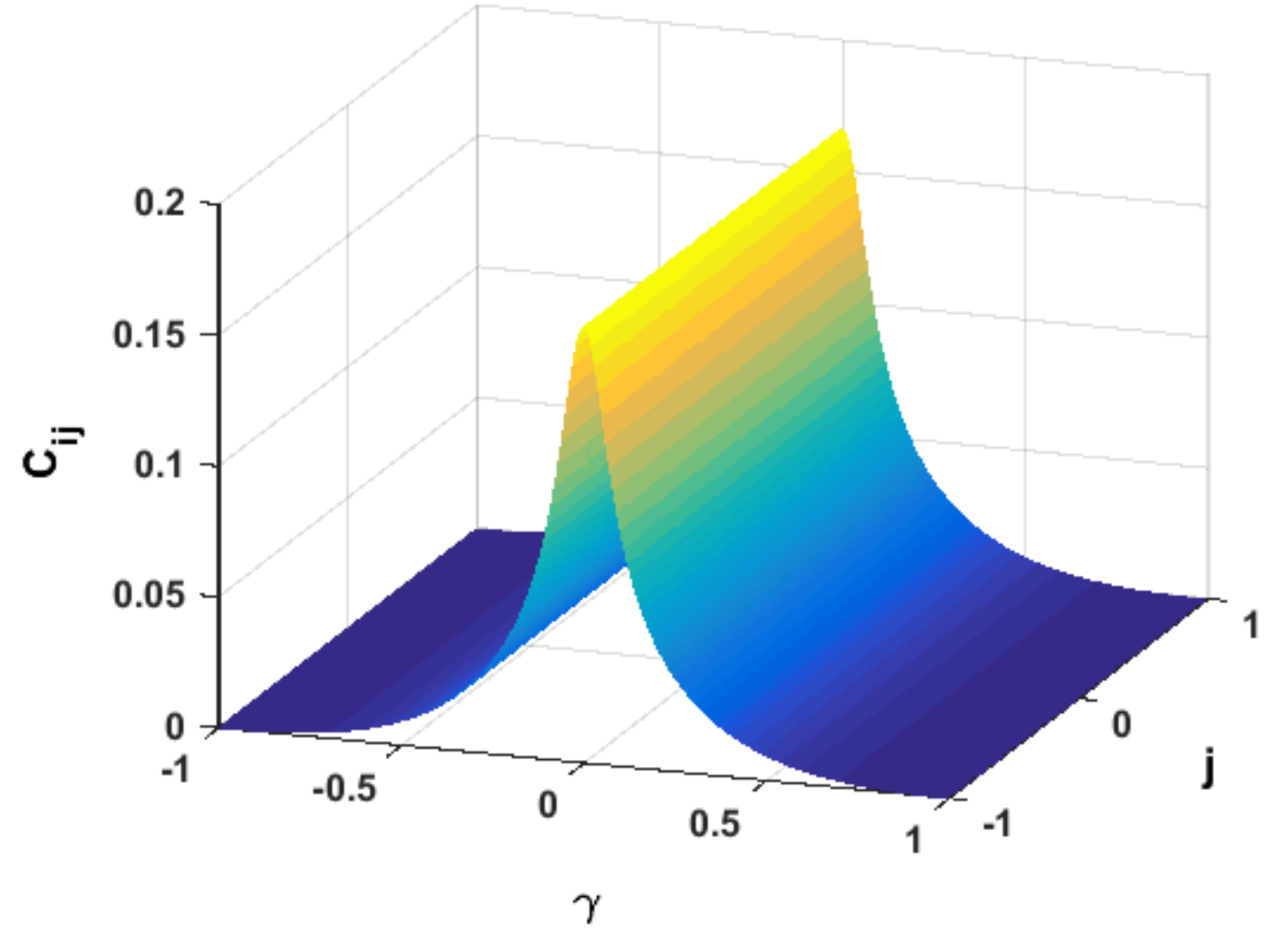}
\caption{\sf Three dimensional plot showing the variation of the concurrence $C_{ij}$ with the control 
parameters $\gamma$ and $j$ for the $0^{th}$ order RG step in the 3-D $XY$ spin-1/2 model. The behavior of the concurrence shows that it is 
independent of the value of the coupling parameter $j$.}
\label{Fig.7.}
\end{figure}


\begin{thebibliography}{99}

\bibitem{nc} M. A. Nielsen and I. L. Chuang, \textit{Quantum Computation and Quantum
Communication} (Cambridge University Press, Cambridge, 2000).

\bibitem{science1} DiVincenzo, David P. Science 270.5234 (Oct 13, 1995): 255.

\bibitem{rmp1} I. M. Georgescu, S. Ashhab and F. Nori, Rev. Mod. Phys. \textbf{86}, 153
(2014).

\bibitem{rmp2} A. Galindo and M. A. Martin-Delgado, Rev. Mod. Phys. \textbf{74}, 347
(2002).

\bibitem{ising}M. Kargarian, R. Jafari, and A. Langari, Phys. Rev. A \textbf{76}, 060304(R)
(2007).

\bibitem{xyz}A. Langari, Phys. Rev. B \textbf{69}, 100402(R)
(2004)

\bibitem{xxz} M. Kargarian, R. Jafari, and A. Langari, Phys. Rev. A \textbf{77}, 032346
(2008).

\bibitem{xxzdm} M. Kargarian, R. Jafari, and A. Langari, Phys. Rev. A \textbf{79}, 042319
(2009).

\bibitem{xxznn} R. Jafari and A. Langari, Physica A \textbf{364}, 213 (2006).

\bibitem{xy} Fu-Wu Ma, Sheng-Xin Liu, and Xiang-Mu Kong, Phys. Rev. A \textbf{83}, 062309
(2011).

\bibitem{xydm} Fu-Wu Ma, Sheng-Xin Liu, and Xiang-Mu Kong, Phys. Rev. A \textbf{84}, 042302
(2011).

\bibitem{fisher} X. M. Liu, W. W. Cheng and J. -M. Liu, Sci. Rep. \textbf{6}, 19359
(2016).

\bibitem{xxzglobal} Sun, W., Shi, J., Wang, D. et al. Quantum Inf Process (2016) \textbf{15}: 245

\bibitem{vedral1} Sylju\aa sen, O., 2003a, Phys. Lett. A \textbf{322}, 25 (2004).

\bibitem{vedral2} Sylju\aa sen, O., 2003b, Phys. Rev. A \textbf{68}, 060301 (R) (2003).


\bibitem{vidal1} G. Evenbly and G. Vidal, Phys. Rev. Lett. \textbf{102, }180406 (2009).

\bibitem{vidal2} G. Vidal, Phys. Rev. Lett. \textbf{99}, 220405 (2007).

\bibitem{2dising} Y. Xu, X. Kong, Z. Liu and C. Wang, Physica A \textbf{446}, 217 
(2016).

\bibitem{Amico-2008} L. Amico, R. Fazio, A. Osterloh, and V. Vedral, Rev. Mod. Phys. \textbf{80},
517 (2008).

\bibitem{wilson1} K. G. Wilson, Rev. Mod. Phys. \textbf{47}, 773 (1975).

\bibitem{dmrgrmp} U. Schollw\"{o}ck, Rev. Mod. Phys. \textbf{77}, 259
(2005).

\bibitem{njp1} S. Gu, G. Tian and H. Lin, New J. Phys. \textbf{8}, 61 (2006).

\bibitem{white1} S. R. White, Phys. Rev. Lett. \textbf{69}, 2863 (1992).

\bibitem{white2} S. R. White and R. M. Noack, Phys. Rev. Lett. \textbf{68}, 3487 (1992).

\bibitem{white3} S. R. White, Phys. Rev. B \textbf{48}, 10345 (1993).


\bibitem{dmrgqit} O. Legeza, R. M. Noack, J. Solyom, and L. Tincani, Applications of 
Quantum Information in the Density-Matrix Renormalization Group, Vol. 739 (Springer, Berlin, 2008).

\bibitem{osborne1} T. J. Osborne and M. A. Neilsen, Quantum Inf Process (2002) \textbf{1}: 45 

\bibitem{sandvik} A. W. Sandvik, AIP Conference Proceedings \textbf{1297}, 135 (2010).

\bibitem{vidal3} G. Vidal, J. I. Latorre, E.Rico and A. Kitaev, Phys. Rev. Lett. \textbf{90}, 227902 (2003).


\bibitem{kadanoff} J. Gonz\={a}lez, M. A. Martin Deigado, G. Sierra and A. H. Vozmediano,
Quantum Electron Liquid and High-T$_{c}$ Super Conductivity, edited by H.
Araki \textit{et al.,} Lecture Notes in Physics Vo38. (Springer, Berlin,
1995), Chap. 11.

\bibitem{xy2d} M. Usman, Asif Ilyas and Khalid Khan, Phys. Rev. A \textbf{92}, 032327
(2015)

\bibitem{osterloh} A. Osterloh, L. Amico, G. Falci and R. Fazio, Nature (London) \textbf{416}, 608 (2002).

\bibitem{isingxy} M. Zhong and P. Tong, J. Phys. A: Math. Theor. \textbf{43}, 505302
(2010).

\bibitem{xxzddim} Shi-Jian Gu, Guang-Shan Tian and Hai-Qing Lin, Phys. Rev. A \textbf{71}, 052322 (2005)


\bibitem{xy26} G. Sierra and M. A. Martin-Delgado, in Strongly Correlated Magnetic and
Superconducting Systems, edited by H. Araki et al., Lecture Notes in Physics
Vol. 478 (Springer, Berlin, 1997).

\bibitem{xy27} M. A. Martin-Delgado and G. Sierra, Int. J. Mod. Phys. A \textbf{11}, 3145 (1996).

\bibitem{wootters} W. K. Wootters, Phys. Rev. Lett. \textbf{80}, 2245 (1998).

\bibitem{osborne2} T. J. Osborne and M. A. Neilsen, Phys. Rev. A \textbf{66}, 032110 (2002)

\bibitem{sachdev} S. Sachdev, Quantum Phase Transitions (Cambridge University
Press, Cambridge, 1999).

\end{thebibliography}
\end{document}